\documentclass[
superscriptaddress,
amsmath,amssymb,
aps,
rmp,
]{revtex4-1}

\usepackage{graphicx}\usepackage{dcolumn}\usepackage{bm}\usepackage{amsmath}
\usepackage[T1]{fontenc}
\usepackage{times}
\usepackage{lmodern}
\usepackage{amssymb}
\usepackage{bm}
\usepackage{graphicx}
\usepackage{gensymb}
\usepackage{mymacros}
\usepackage{mathtools}
\usepackage{float}
\usepackage{dsfont}
\usepackage{romannum}
\usepackage{multirow}
\usepackage{amsthm}
\usepackage{colortbl}
\usepackage{subfig}
\usepackage{xr}
\usepackage{appendix}
\usepackage{textcomp,libertine}
\usepackage{hyperref}
\usepackage[mathlines]{lineno}

\usepackage[
margin=0.5in,
]{geometry}

\begin{document}

\title{
High-energy coherent X-ray diffraction microscopy of polycrystal grains: \\
first steps towards a multi-scale approach}

\author{S. Maddali}
\email{smaddali@anl.gov}
\affiliation{Materials Science Division, Argonne National Laboratory, Lemont, IL 60439 (USA)}

\author{J.- S. Park}
\affiliation{X-ray Sciences Division, Argonne National Laboratory, Lemont, IL 60439 (USA)}\author{H. Sharma}
\affiliation{X-ray Sciences Division, Argonne National Laboratory, Lemont, IL 60439 (USA)}\author{S. Shastri}
\affiliation{X-ray Sciences Division, Argonne National Laboratory, Lemont, IL 60439 (USA)}\author{P. Kenesei}
\affiliation{X-ray Sciences Division, Argonne National Laboratory, Lemont, IL 60439 (USA)}\author{J. Almer}
\affiliation{X-ray Sciences Division, Argonne National Laboratory, Lemont, IL 60439 (USA)}\author{R. Harder}
\affiliation{X-ray Sciences Division, Argonne National Laboratory, Lemont, IL 60439 (USA)}

\author{M. J. Highland}
\affiliation{Materials Science Division, Argonne National Laboratory, Lemont, IL 60439 (USA)}

\author{Y. Nashed}
\affiliation{Mathematics and Computer Science Division, Argonne National Laboratory, Lemont, IL 60439 (USA)}

\author{S. O. Hruszkewycz}\affiliation{Materials Science Division, Argonne National Laboratory, Lemont, IL 60439 (USA)}

\begin{abstract}
	We present proof-of-concept imaging measurements of a polycrystalline material that integrate the elements of conventional high-energy X-ray diffraction microscopy with coherent diffraction imaging techniques, and that can enable in-situ strain-sensitive imaging of lattice structure in ensembles of deeply embedded crystals over five decades of length scale upon full realization. 
Such multi-scale imaging capabilities are critical to addressing important questions in a variety of research areas such as materials science and engineering, chemistry, and solid state physics. 
Towards this eventual goal, the following key aspects are demonstrated: 1) high-energy Bragg coherent diffraction imaging (HE-BCDI) of sub-micron-scale crystallites at 52 keV at current third-generation synchrotron light sources, 2) HE-BCDI performed in conjunction with far-field high-energy diffraction microscopy (ff-HEDM) on the grains of a polycrystalline sample in a smoothly integrated manner, and 3) the orientation information of an ensemble of grains obtained via ff-HEDM used to perform complementary HE-BCDI on multiple Bragg reflections of a single targeted grain. 
These steps lay the foundation for integration of HE-BCDI, which typically provides a spatial resolution tens of nanometers, into a broad suite of well-established HEDM methods, extending HEDM beyond the few-micrometer resolution bound and into the nanoscale, and positioning the approach to take full advantage of the orders-of-magnitude improvement of X-ray coherence expected at fourth generation light sources presently being built and commissioned worldwide.

 \end{abstract}
\maketitle

\section{Introduction}
\label{S:intro}
The ability to non-destructively measure local lattice distortion effects in crystalline materials with X-ray microscopy enables connections to be made between nano- and micro-scale structure with macroscopic properties under working conditions, thus playing an important role in advancing materials science and engineering, chemistry, and solid state physics. 
For example, recent successful studies include characterizing electro-catalytic processes \emph{in situ} \cite{Ulvestad2015,Ulvestad2015a}, informing the design of structural, functional, and quantum materials \cite{Hruszkewycz2012,Highland2017,Hruszkewycz2018}, and testing and validating theoretical models of polycrystals under various real-world stimuli\cite{Suter2008,Schuren2015}.
Radiation of different energies in the hard X-ray regime is employed for a variety of such diffraction-based measurements at synchrotron light sources.
At the lower end of this energy range (7-15 keV) where coherent illumination is readily achieved, scattering methods such as Bragg coherent diffractive imaging (BCDI) and Bragg ptychography allow the measurement of the local lattice strain in single crystals with nanoscale spatial resolution \cite{Robinson2001,Miao2015,Robinson2009,Hruszkewycz2012,Hruszkewycz2017a}, and such approaches have recently been demonstrated for imaging individual grains in a polycrystal \cite{Yau2017}.
However, because the penetration length of these X-ray photons is typically a few tens of micrometers in dense solids, coherent diffraction imaging methods cannot be used to interrogate grains deep within polycrystalline bulks beyond measurement of surface-facing grains ~\cite{Vaxelaire2014,Yau2017,Cherukara2018}.
At the other end of the hard X-ray energy range (higher than 50 keV), high-energy diffraction microscopy (HEDM) techniques that do not rely on beam coherence permit the imaging and volume-averaged strain characterization of thousands of grains in a polycrystalline bulk of up to a millimeter in size\cite{Suter2006,Bernier2011}.
Querying such large samples in this manner becomes possible owing to 
the higher penetration at high X-ray energies, and the current state of the art in HEDM allows the average strain state, lattice orientation, unit cell structure, and morphology of individual grains that make up the polycrystal to be determined at a spatial resolution of about $1.5~\mu$m~\cite{Bernier2011}.

Presently HEDM and BCDI have their respective places in materials research, 
and combining the two techniques into a single multi-component microscopy approach presents a tremendous new opportunity.
A composite measurement scheme that draws from the strengths of both methods could potentially open the door to a new range of possibilities in materials characterization by enabling strain imaging of bulk materials with highly intricate crystallographies and domain morphologies, resolved from millimeters to nanometers.
Measurements of this kind will enhance our understanding of grain and interface dynamics in solids by enabling access to length scales that elucidate a variety of lattice distortion features (such as macroscopic stress concentration sites down to individual dislocations), and would present immensely valuable insights.
One important example is that of the mechanical behavior of bulk polycrystalline structural materials, for which fully descriptive models of deformation and failure have not been fully realized despite decades of study~\cite{Kapoor2018}. 
Further progress towards this and other important materials questions depends on \emph{in situ} experimental methods that can measure structural processes over many length scales, as could be realized by combining the HEDM and BCDI modalities.

Several critical steps towards the unification of BCDI and HEDM are demonstrated in this article, anticipating the capabilities of fourth-generation light sources coming online worldwide now and in the near future~\cite{Barber2014}. 
These new synchrotron light sources will provide a several-hundred-fold increase in coherent X-ray flux as compared to today's facilities, making high-energy BCDI practical and routine.
In this context, a high-throughput materials characterization capability that makes use of high X-ray energies and that consists of combined and fully integrated HEDM and BCDI modalities will be realizable. 
We present proofs-of-concept of several elements of such a measurement using X-rays with an energy $52$ keV at the Advanced Photon Source, a third generation 7 GeV synchrotron equipped with a superconducting undulator for enhanced photon flux~\cite{Ivanyushenkov2017}. 
Several key aspects are established:
\begin{enumerate}
    \item With the appropriate adaptation of HEDM X-ray optics, it is possible to implement high-energy BCDI (HE-BCDI) on an isolated sub-micron-scale crystal at 52 keV, provided partial coherence effects of the X-ray beam are accounted for.  
    \item It is possible to integrate far-field HEDM (ff-HEDM) and HE-BCDI and to make these measurements in succession on individual grains in a polycrystalline material. 
    \item  In a polycrystal, the orientation information of each grain within a broad illumination footprint that is obtained from ff-HEDM can be used to efficiently perform HE-BCDI on multiple Bragg reflections of a single grain within a wide window of reciprocal space, a capability that is extremely difficult to realize in a standalone BCDI measurement. 
\end{enumerate}

These demonstrations, detailed below, lay the foundation for a flexible new multi-modal means of studying polycrystalline materials in situations where the long penetration depths of high energy X-rays is critical. 
 
\section{Demonstration of HE-BCDI}
\label{SS.hebcdidemo}
\begin{figure*}
	\centering
	\includegraphics[width=0.75\textwidth]{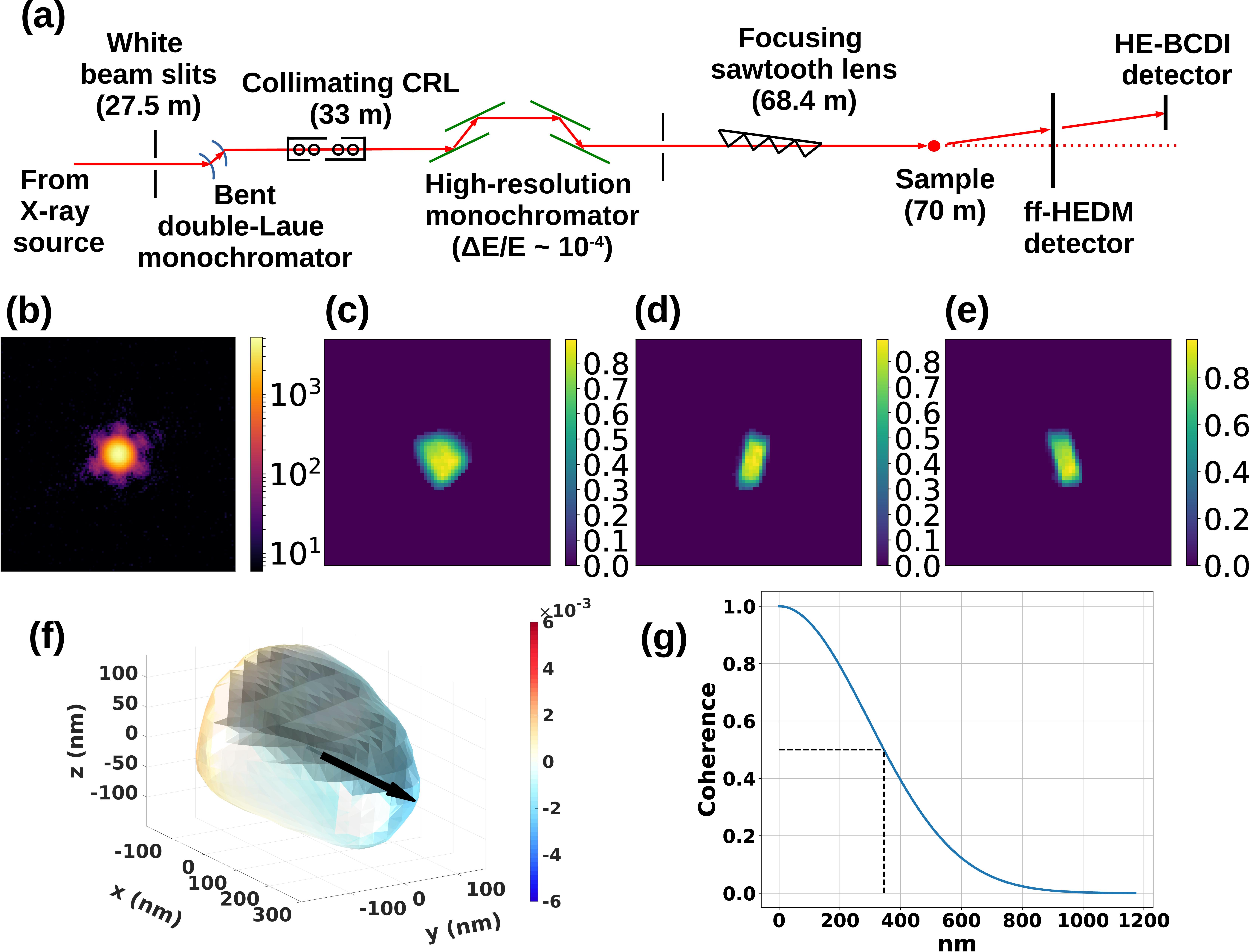}
	\caption{
		\textbf{(a)} Simplified schematic of Beamline 1-ID-E at the Advanced Photon Source, showing the essential components. Only the top sawtooth lens was used to provide focus. 
		See Supplementary Material, Section 1 for details on why this configuration was chosen. 
		\textbf{(b)} Central slice of the partially coherent 3D diffraction signal from the gold nanoparticle.
		\textbf{(c)-(e)} Mutually perpendicular amplitude cross sections of the gold nanoparticle after phase retrieval.
		\textbf{(f)} Lattice strain profile on the surface of the 3D nanoparticle. 
		The isosurface region with a seemingly sudden drop in strain is actually due to the 3D lighting effects that better highlight the contours of the object.
		The arrow indicates the direction of horizontal transverse coherence of the X-ray beam (laboratory $X$).
		\textbf{(g)} Profile of the Gaussian coherence function along the direction of the arrow shown in (f).
		The half-width at half maximum  is an estimate of the beam coherence length along the direction of interest.
		}
	\label{fig:expSchematic}
\end{figure*}

The ability to perform HE-BCDI measurements with a partially coherent high-energy beam is a key requirement for the eventual integration of the HEDM and BCDI measurement modalities.
To this end, one of the goals of our study was to develop an experimental configuration suitable for HE-BCDI at a third generation synchrotron, by imaging an isolated nanocrystal obtained by de-wetting a gold film on a silicon substrate.
A simplified schematic of the X-ray optical configuration used for this purpose is shown in Fig.~\ref{fig:expSchematic}(a), which yielded a coherence volume and detector geometry suitable for HE-BCDI imaging at the 1-ID-E end station of the Advanced Photon Source (see Supplementary Material, Section 1). 
The critical components needed to realize this were white beam slits opened to $0.5$ mm $\times~0.5$ mm, a collimating lens, a high-resolution monochromator~\cite{Shastri2004}, a vertically-focusing sawtooth lens~\cite{Shastri2007}, a sample-detector distance of $6.12$ m, and a finely pixelated photon-counting area detector with good quantum efficiency at 52 keV (Lynx 1800 detector manufactured by Amsterdam Scientific Instruments, with a Medipix3 chip and a 55 $\mu$m pixel pitch). 
With this  arrangement, a $1.5 \times 50$ $\mu$m beam was delivered to the sample. 
The detector was mounted along the far wall of the experimental enclosure with the two degrees of freedom (radial distance $R$ and azimuthal angle $\eta$) needed to reach Bragg peaks with a scattering angle of $\sim 15^\circ$. 

With this arrangement, the three-dimensional diffraction pattern in the vicinity of a $\left\langle 111 \right\rangle$ Bragg peak from the gold nanoparticle was measured using the high-precision sample manipulation system at the 1-ID-E end station~\cite{Benda2016}.
The sample was oriented such that the diffracted exit beam passed through the silicon substrate en route to the detector. 
The rotation angle $\omega$ of the sample was incremented about the vertical axis in fine steps of $\Delta \omega = 0.01^\circ$ spanning a total of $0.6^\circ$ about the angle of the Bragg peak maximum. At each sample angle, area detector images of the diffraction patterns were collected with an exposure time of 600 seconds per angle. 
The three-dimensional intensity data acquired in this manner encoded the morphology and the lattice distortions of the diffracting crystal.

A 2D slice of the diffraction around the Bragg peak is seen in Figure~\ref{fig:expSchematic}(b), in which the fringes are oversampled (as typically required in BCDI), but showing that the fringe contrast is less than 100\%, indicating partially coherent illumination.
The 3D image of the scatterer was reconstructed using a BCDI phase retrieval approach that accounts for partial coherence, based on recently published methods ~\cite{Tran2005,Clark2011,Clark2012}. 
Within the phase retrieval algorithms used, the partial coherence correction was achieved by modeling the measured diffraction as a convolution of the fully coherent diffraction intensity pattern with a blurring kernel~\cite{Tran2005,Clark2011,Clark2012}, which was chosen to be a multivariate Gaussian.
The phase retrieval recipe consisted of alternating cycles of the Gerchberg-Saxton error-reduction and hybrid input-output~\cite{Fienup1982} along with intermittent updating of the real-space support \emph{via} a shrinkwrap algorithm~\cite{Marchesini2003} and optimization of the six parameters of the unknown Gaussian blurring kernel (see Supplementary Material, Section 2 for the exact phase retrieval algorithm and kernel parameterization).

This approach resulted in a complex-valued real-space image of the diffracting crystal $\boldsymbol{\rho}$ that encodes information about lattice displacement distribution within the particle. 
Specifically, the phase of $\boldsymbol{\rho}$ is related to the projection of the spatial distribution of atomic lattice displacement perturbations $\mathbf{u}$ along a reciprocal lattice vector $\mathbf{Q}$~\cite{Robinson2009}, such that: $\boldsymbol{\rho} = \left|\boldsymbol{\rho}\right| \exp\left(2\pi i \mathbf{u} \cdot \mathbf{Q}\right)$.
Via the phase of $\boldsymbol{\rho}$ (\emph{i.e.} $\angle{\boldsymbol{\rho}}$), one can obtain a component of the displacement field: $u_{111} = \angle{\boldsymbol{\rho}} / | 2\pi \mathbf{Q}_{111} |$.
Subsequently, the component of the strain tensor along $\mathbf{Q}_{111}$ can be determined at a location $x_{111}$, by computing the partial derivative $\partial u_{111}/\partial x_{111}$.

Figure~\ref{fig:expSchematic}(c)-(e) denote mutually perpendicular amplitude cross-sections of the reconstructed nanoparticle (\emph{i.e.} $\left|\boldsymbol{\rho}\right|$), obtained from phase retrieval. 
Figure~\ref{fig:expSchematic}(f) shows a 3D image of the particle with the color scale representing the surface variations of the strain component $\partial u_{111}/\partial x_{111}$.
The particle itself has a maximum diameter of about 400 nm, displays distinct facets, and has relatively low levels of strain -- all characteristics typical of gold particles obtained by dewetting~\cite{Cha2016}. 

The arrow in Figure~\ref{fig:expSchematic}(f) indicates the horizontal ($X$-) direction in the laboratory frame, and the profile of the real-space representation of the blurring kernel along this direction is shown in Figure~\ref{fig:expSchematic}(g).
We find that in this direction the half-width at half-maximum of the Gaussian kernel is about 350 nm. This length is a rough estimate of the 50\% coherence threshold in that dimension.

This result establishes a baseline for the HE-BCDI methodology and suggests that with the X-ray optical setup employed in this work, crystallites with diameters corresponding roughly to the 50\%-coherence threshold of the beam can be imaged with HE-BCDI. Crucially, this HE-BCDI reconstruction of a symmetrically faceted, relatively strain-free nanoparticle provided a sufficiently well-constrained measurement of the blurring kernel that could be used to model the partial coherence of the X-ray beam for subsequent HE-BCDI measurements of individual grains that in general displayed more disordered fringe patterns owing to potentially greater extents of lattice strain and less regular faceting.

\section{Integrated ff-HEDM and HE-BCDI measurements}
\label{SS.hedmbcdigrain}
\begin{figure}
    \centering
    \includegraphics[width=0.5\textwidth]{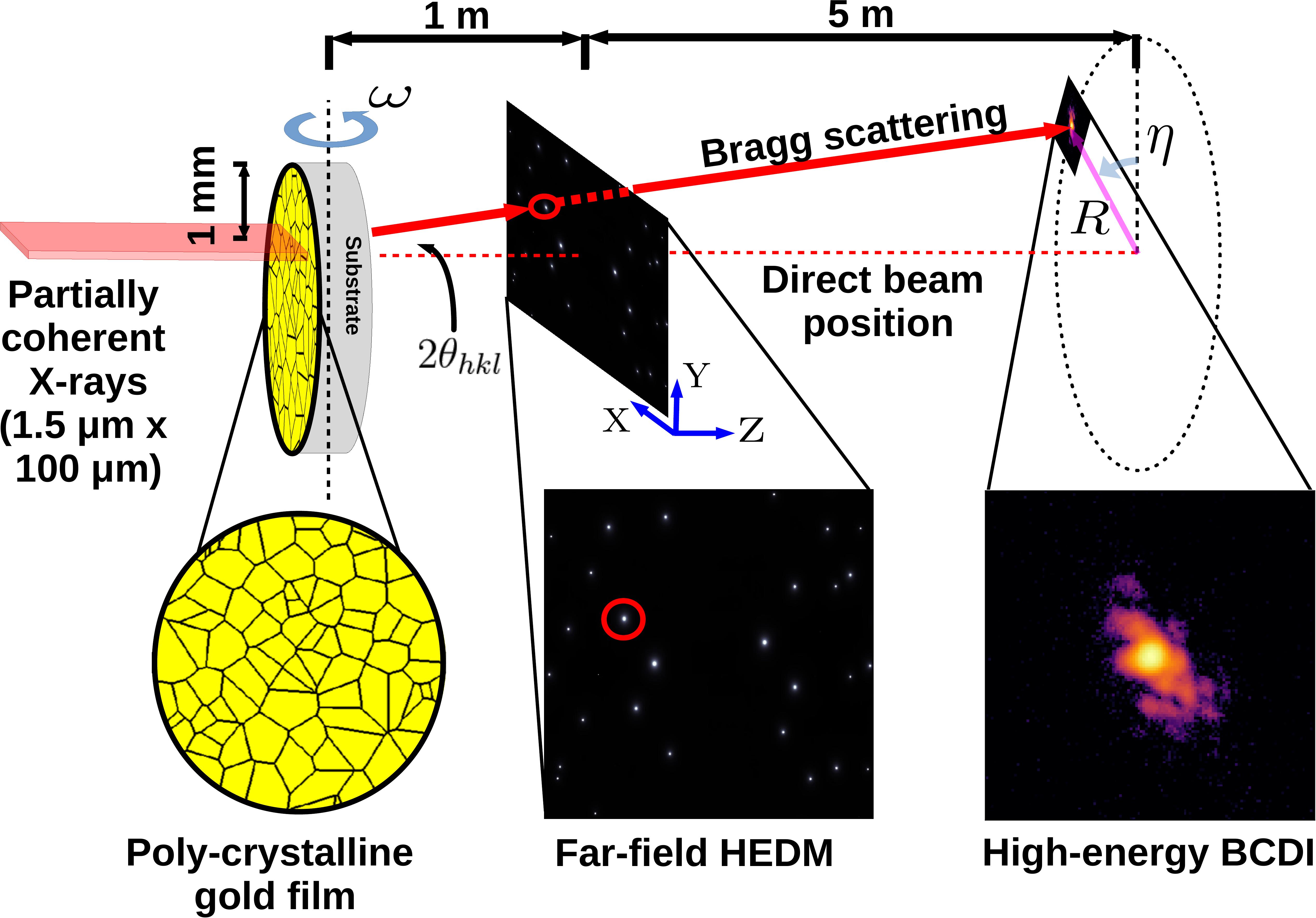}
    \caption{
    Schematic of the multi-scale measurement technique with the HEDM and BCDI components. 
    The ff-HEDM measurement is made at a scattering distance of $1$ m while the HE-BCDI measurement is made at a scattering distance of $6$ m.
    The detector images respectively denote the relatively coarse resolution of the Bragg peaks from the grains in the sample (ff-HEDM), and the finer details of the fringe pattern in the vicinity of one such peak.
    We note that the beam dimensions shown here only correspond to the ff-HEDM measurement.
    The beam size was actually set to $1.5  \times 50 \mu$m for all HE-BCDI measurements.
    }
    \label{fig:HEDMBCDI}
\end{figure}

\begin{figure}
    \centering
    \includegraphics[width=0.5\textwidth]{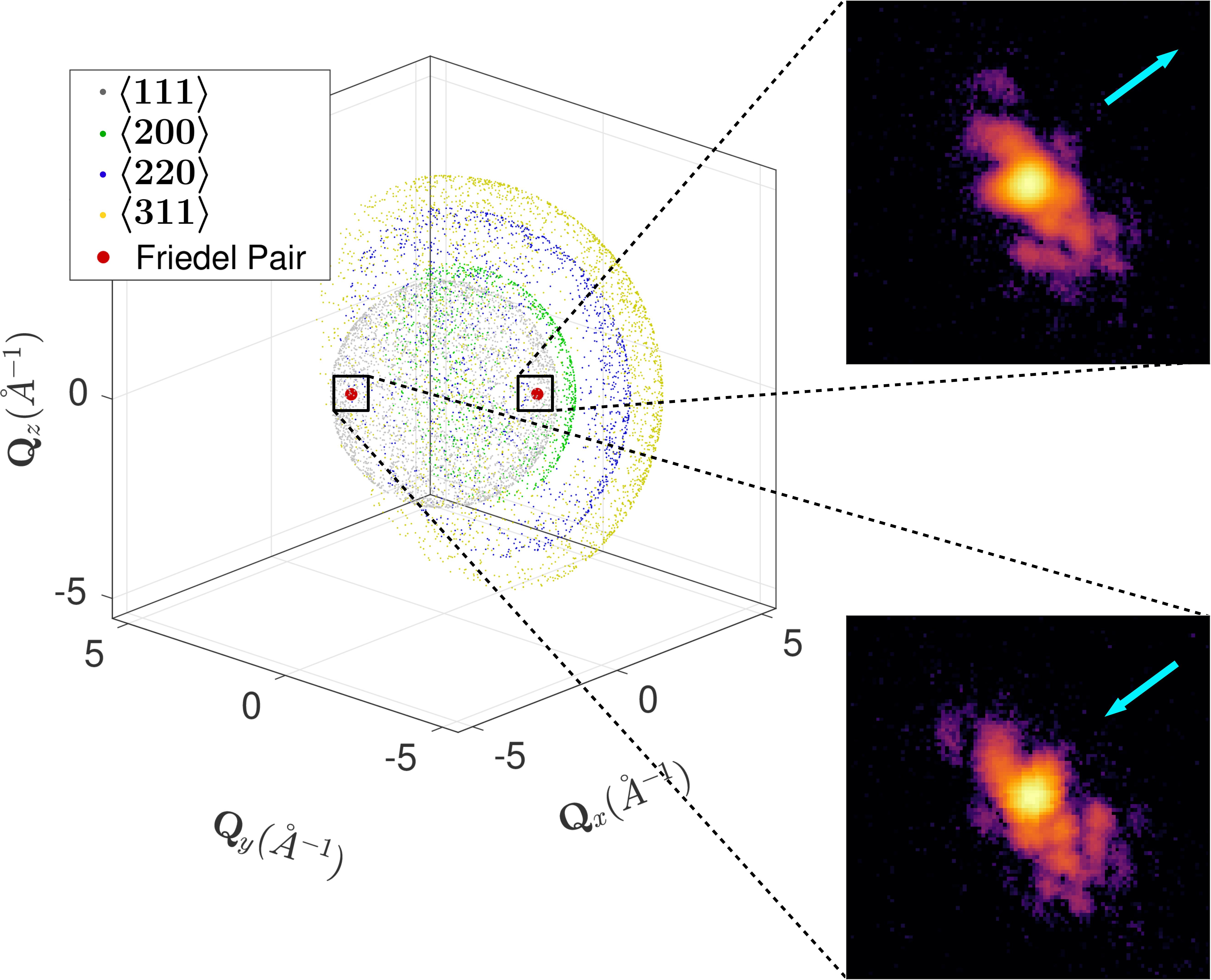}
    \caption{
    Scatter plot of the reciprocal lattice points corresponding to the Bragg peaks from the illuminated grains, acquired during the ff-HEDM measurement.
    Shown here is a subset of the observed Bragg reflections, up to the $\left\langle 311 \right\rangle$ reflection for fcc crystals.
    The bold markers denote a Friedel pair of $\left\langle 111 \right\rangle$ reflections from a single grain.
    Also shown are the ``zoomed-in'' central slices of the measured partially coherent diffraction signal in the vicinity of these reflections (reoriented to emphasize the centrosymmetry about the line denoted by the two arrows, corresponding to reciprocal lattice vector directions $\pm \mathbf{Q}_{111}$ respectively).
    }
    \label{fig:zoomins}
\end{figure}

To demonstrate the integration of the HE-BCDI measurement modality described above with standard HEDM measurements, we utilized a polycrystalline gold film~\cite{Yau2017} which was deposited by electron beam evaporation on an amorphous carbon substrate. 
The film thickness was around 300 nm, with a characteristic in-plane grain size of 400 nm. 
A schematic showing the different detectors used in the combined ff-HEDM and HE-BCDI experiment is shown in Figure~\ref{fig:HEDMBCDI}.
The sample was oriented with the substrate surface initially normal to the incident beam, and first measured with ff-HEDM. 
The line-focused X-ray beam obtained with the vertical saw-tooth focusing lens had a footprint on the sample of $1.5 \times 100$ $\mu$m, with an X-ray energy of 52 keV, as before. 
The ff-HEDM measurement was performed by rotating the sample about the $Y$ axis through $360^\circ$ using a high-precision Aerotech rotation stage (designated $\omega$ in Figure~\ref{fig:HEDMBCDI}) that scanned the sample in angular increments of $0.01^\circ$, resulting in 36,000 acquired diffraction images. 
Bragg peaks were measured during this scan with a bank of four GE-41RT detectors positioned $\sim 1$ m from the sample with $200 \mu m$ pixels that subtended complete Debye-Sherrer rings out to scattering angles of up to $\sim 15^\circ$. 
Over the course of the entire scan, tens of thousands of diffraction peaks were collected, corresponding to the $\left\langle 111 \right\rangle$, $\left\langle 200 \right\rangle$, $\left\langle 220 \right\rangle$, and $\left\langle 311 \right\rangle$ Bragg reflections of the face-centered cubic gold lattice. 
A visualization of a subset of these measured Bragg peaks is shown in Figure~\ref{fig:zoomins}, in which the distinct Debye spheres of each order of Bragg diffraction from the illuminated grains are denoted by differently colored markers.  

The HEDM software suite MIDAS~\cite{Sharma2012,Sharma2012a} was used to automatically identify and index all of the peaks in the measured detector images and to map each peak back to one of the 6768 grains within the illuminated beam volume. The details of this indexing process  are given in Ref.~\cite{Bernier2011}, and is standard procedure for ff-HEDM experiments. 
The quantities $\omega$ and $\eta$ were determined up to an uncertainty of $0.01^\circ$. 
This provided angular information of sufficient precision in order to orient the sample and the ASI detector at 6.12 m from the sample for a HE-BCDI measurement from a chosen grain in the polycrystal (described in the next section). 

We note that both ff-HEDM and HE-BCDI measurements were made sequentially with the same arrangement of X-ray optics and goniometer hardware, differing only in the sample-detector distance and the detector module used. 
This demonstration points towards the potential of smooth physical integration of ff-HEDM and HE-BCDI, and also indicates the possibility of integration with other high energy microscopy imaging modes, including near field HEDM (nf-HEDM)~\cite{Suter2006} and diffraction contrast tomography (DCT)~\cite{Ludwig2008}. 
Physically realizing this unification requires significant development that is currently ongoing, both in terms of improving the spatial resolution of near-field HEDM detectors and line-focusing x-ray optics to reach sub-micrometer length scales (thereby overlapping with the demonstrated length regime of HE-BCDI), and in terms of improving synchrotron storage ring technology that will enable HE-BCDI of ten-micron-scale grains (thus intersecting with current nf-HEDM capabilities).

\section{HE-BCDI of a Bragg peak Friedel pair from a single grain}
\label{SS:grainhebcdi}
The grain orientation and Bragg peak indexing information from the ff-HEDM measurement was used to measure and image two different $\left\langle 111\right\rangle$ Bragg peaks from the same grain with HE-BCDI. Specifically, we sought to demonstrate unambiguously that the selected peaks indeed originated from the same grain by reconstructing HE-BCDI data from a Friedel pair of Bragg peaks ($\left[111\right]$ and $\left[\bar{1} \bar{1} \bar{1}\right]$) which are centrosymmetric about the origin of reciprocal space.
Since Friedel pairs encode equivalent structural information due to their centrosymmetry (see Supplementary Material, section 3), the subsequent HE-BCDI reconstructions were expected to show equivalent strain profiles.
The absolute reciprocal space positions of the chosen Friedel pair reflections (obtained via ff-HEDM) are shown in Figure~\ref{fig:zoomins}, along with 2D images of the fringe detail about each of these peaks, obtained with the HE-BCDI detector. 
It is clear that the Friedel pair of peaks are located in centrosymmetric positions about the reciprocal space origin $\mathbf{Q} = 0$, and that the ``zoomed-in'' view of these peaks provided by the HE-BCDI detector images also shows consistent centrosymmetric fringe patterns.

This pair of peaks was chosen in order to ensure successful HE-BCDI reconstructions based on the fact that the diffracted signal was relatively strong from the originating grain, and the volume of reciprocal space in the vicinity of the peak did not overlap with Bragg peaks from other grains. 
The HE-BCDI measurements were similar to the case of the isolated gold nanoparticle ($60$ angular steps of size $\Delta \omega = 0.01^\circ$, each with an exposure of $600$ seconds). 
Image reconstruction was done using the same phase retrieval approach as above, accounting for partial coherence by utilizing the Gaussian blurring kernel determined from the isolated crystal as a starting guess and allowing further minor refinement of the Gaussian parameters during phase retrieval. 

\begin{figure}
    \centering
    \includegraphics[width=0.5\textwidth]{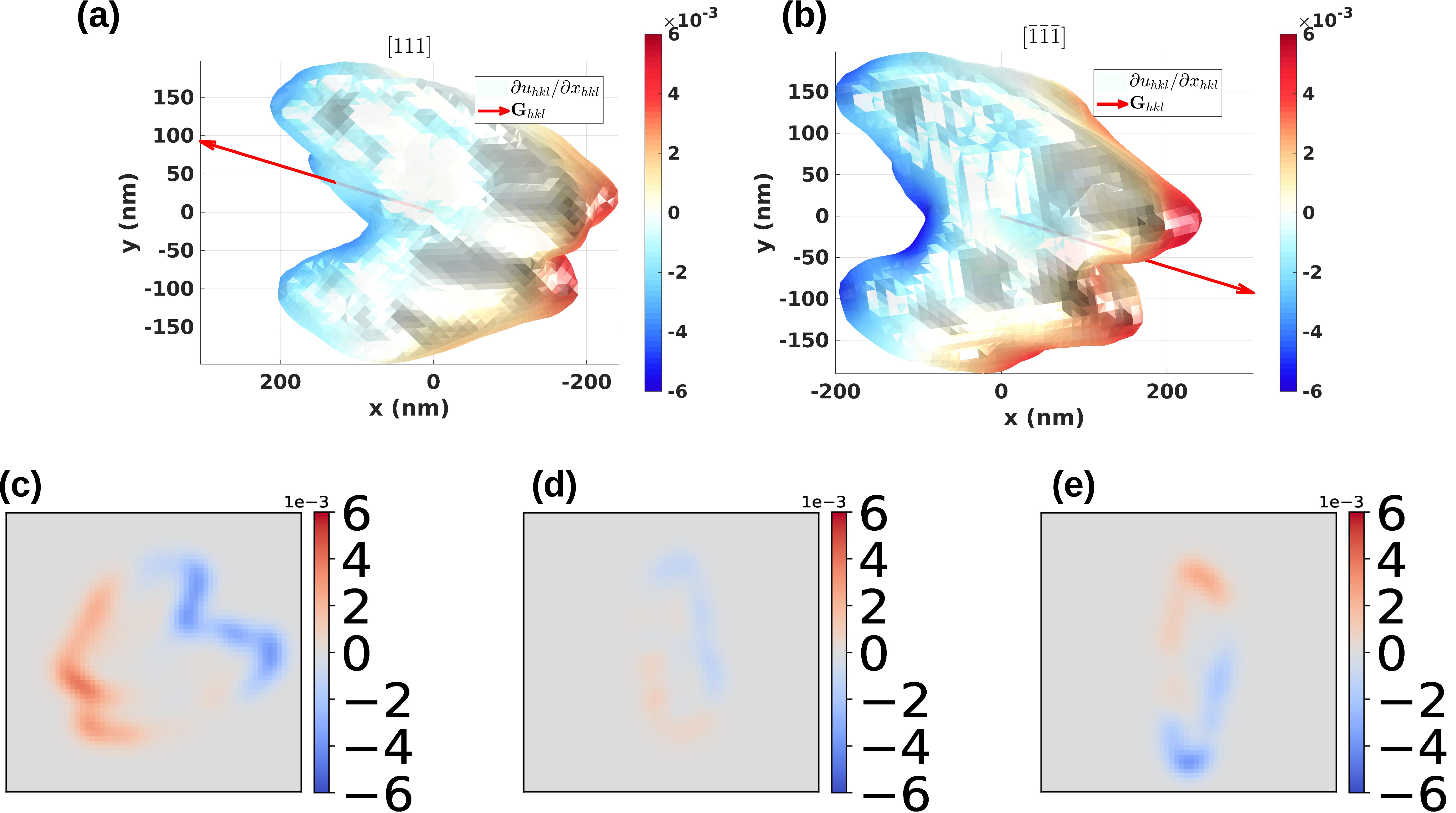}
    \caption{
        \textbf{(a)} Reconstructed image of the grain of interest from the polycrystalline film, obtained from the HE-BCDI data set corresponding to the $\left[111\right]$ reflection. 
		The associated strain field component is superposed on its surface.
        \textbf{(b)} The same grain imaged from the data set corresponding to the $\left[\bar{1}\bar{1}\bar{1}\right]$ reflection.
		As with Figure~\ref{fig:expSchematic}(f), the regions of seemingly sudden drops in strain are actually due to 3D lighting effects.
\textbf{(c)-(e)} Strain profile cross-sections of the grain reconstruction shown in (a), through the center of the grain.
    }
    \label{fig:strain_cs}
\end{figure}

The 3D image reconstructions from both centrosymmetric Bragg peaks are shown in Figure~\ref{fig:strain_cs} with surface-strain coloration. Also shown are cuts through the interior of the grain that show the spatial distribution of the strain component $\partial u_{111}/\partial x_{111}$.
We see that the center of the grain is relatively strain free, while the regions of relatively significant strain are seen to be along the interfaces with neighboring grains (\emph{i.e.} the grain boundaries). 

In comparing the 3D image reconstructions, we immediately recognize that the
two images are of the same grain,
indicating that the two Bragg peaks that were measured from among tens of thousands emanating from the illuminated sample were indeed a Friedel pair. 
The morphology, orientation, and strain state of the two images are very similar, as expected, and small differences can be ascribed to factors such as the low signal-to-noise ratios in these measurements and to the inherently different sampling of 3D reciprocal space in the two measurements. 
Importantly, the HE-BCDI reconstruction gave access to the lattice distortion in the interior of this grain with an approximate spatial resolution of $47.5$ nm (see Section 4 of the Supplementary material), pointing to the broader potential of extending the HEDM technique (whose spatial resolution is about $1.5~\mu$m) with coherent diffraction imaging. 

This measurement demonstrates the efficacy of using information obtained from ff-HEDM to efficiently execute a complementary HE-BCDI measurement that resolves intra-granular strain fields critical to the behavior and properties of polycrystalline materials.  
Our work described a demonstration involving two
equivalent HE-BCDI images of a single grain obtained from centrosymmetric Bragg peaks. However, the full power of the approach will be the ability to measure many more non-equivalent Bragg peaks from a given grain in order to spatially resolve the full $6$-component strain tensor, as has  been demonstrated for isolated nanocrystals at lower energies ($< 10$ keV)~\cite{Newton2010,Hofmann2017}. 
In this regard, utilizing high energy x-rays is particularly appealing because it presents convenient access to higher-order Bragg peaks (as shown in Figure \ref{fig:zoomins}) which provide higher strain resolution and which are difficult or impossible to access at lower beam energies.
As a step in this direction, 
we present an image reconstruction of another grain in the polycrystal film, obtained from a HE-BCDI measurement from a $\left\langle 200 \right\rangle$ Bragg peak (see Section 5 of the Supplementary Material), that
represents another benchmark in the eventual realization of integrated multi-scale measurements with high-energy coherent X-rays.

\section{Conclusions}
\label{S:conclusions}
We have demonstrated key steps towards the realization of a fully integrated multi-modal high energy diffraction microscopy capability at synchrotron light sources. 
Such a capability will enable fundamentally new in-situ structural imaging studies of polycrystalline materials over four decades of length scale, in deeply embedded environments, and with strain sensitivity as fine as $1 \times 10^{-5}$.  
High-throughput implementation of such an approach, while impractical today due to the very limited coherence of high energy X-rays at today's third generation light sources, will be achievable at fourth generation synchrotrons  coming online in the near future that promise up to three orders of magnitude increase in coherent flux within a wide range of X-ray energies.  
In this context, HE-BCDI measurements accelerated to few-minutes time scales integrated with HEDM will be one of the many possible means by which to capitalize on the much-improved coherence properties of these fourth-generation sources.

Finally, we note that the HE-BCDI measurements envisioned in such future experiments could differ from the approach presented in this paper, which consisted of a standard-practice variety  of BCDI (\emph{i.e.} diffraction pattern oversampling ensured via long sample-to-detector distance, followed by reconstruction with standard algorithms). 
In particular, recent work has been devoted to developing new strategies for HE-BCDI that relax the requirement for the very long sample-to-detector distance by employing novel hardware~\cite{Pedersen2018,Pedersen2018a}, phase retrieval~\cite{Maddali2018a}, and signal processing solutions~\cite{Maddali2018} that will significantly aid in the realization of coherence-aided multi-modal high energy materials microscopy methods.

\section{Acknowledgements}
\label{S:acknowledge}
Conceptualization of the high-energy BCDI experiment and its integration with the far-field HEDM experimental modality, as well as subsequent phase retrieval and data analysis was supported by Laboratory Directed Research and Development (LDRD) funding from Argonne National Laboratory, provided by the Director, Office of Science, of the U.S. Department of Energy under Contract No. DE-AC02-06CH11357. 
The experimental demonstration of the method was supported by the U.S. Department of Energy, Office of Science, Basic Energy Sciences, Materials Science and Engineering Division.
This research uses the resources of the Advanced Photon Source, a U.S. Department of Energy (DOE) Office of Science User Facility operated for the DOE Office of Science by Argonne National Laboratory under Contract No. DE-AC02-06CH11357. 
\section{Author contributions}
\label{S:authcontrib}
The manuscript was written by SM and SOH.
The measurements were carried out by SM, SOH, JSP, HS and PK.
SS developed the high-energy X-ray optics and optimized it for coherent diffraction measurements.
The gold film was fabricated by MJH. 
The gold nano-particle sample was provided by RH. 
The phase retrieval reconstructions and subsequent strain analysis were done by SM with help from SOH, RH, YN, JSP and PK. 
All authors contributed to the refining of the manuscript.

\appendix

\section{Configuration of beamline optics}
\label{A:optics}
\begin{figure}
	\centering
	\includegraphics[width=0.5\textwidth]{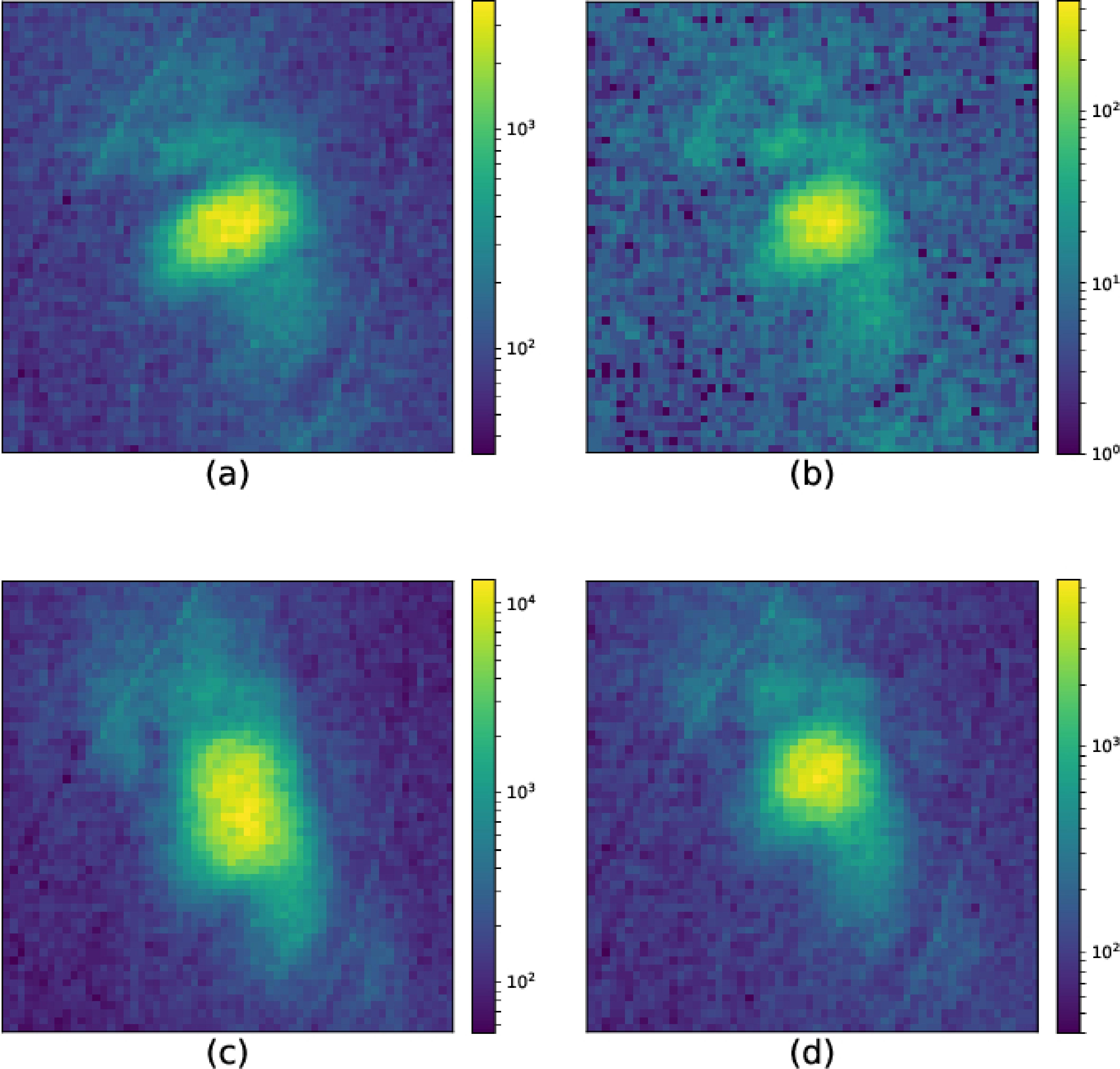}
	\caption{
		The four diffraction spots, taken in order, correspond to configurations 1, 2, 3 and 5 from Table~\ref{tab:optics} respectively.
		}
		\label{fig:S:optics}
\end{figure}

The ff-HEDM and HE-BCDI measurements were made at Beamline 1-ID-E of the Advanced Photon Source.
We refer to the simplified beamline schematic in Fig.1(a) of the main manuscript. 
Determining the optimal configuration for the optics for the HE-BCDI measurements involved trying various combinations of the high-resolution monochromator (HRM), the collimating CRL and the two vertical focusing sawtooth lenses in the X-ray beam.
The configurations were tested by a visual estimate of the sharpness of the partially coherent diffraction pattern around a suitable isolated Bragg reflection from the polycrystalline gold thin film. 
The spot of interest was observed on a photon-counting detector at a sample-detector distance of $6.12$ m and a beam energy of $52$ keV.
The various combinations are summarized in Table~\ref{tab:optics} and the observed diffraction spot corresponding to four of these configurations are shown in Fig.~\ref{fig:S:optics}. 
\begin{table}[htbp]
	\caption{Various configurations of beam-line optics}
	\begin{tabular}{|c|c|c|c|c|}
		\hline
		\multicolumn{ 1}{|c|}{\textbf{Configuration}} & \multicolumn{ 4}{c|}{\textbf{Component in X-ray beam}} \\ \cline{ 2- 5}
		\multicolumn{ 1}{|c|}{} & \textbf{HRM} & \textbf{Collimator} & \textbf{Top focusing} & \textbf{Bottom focusing} \\ \hline
		1 & No & No & No & No \\ \hline
		2 & Yes & No & No & No \\ \hline
		3 & Yes & Yes & Yes & Yes \\ \hline
		4 & Yes & Yes & No & No \\ \hline
		5 & Yes & Yes & Yes & No \\ \hline
		6 & Yes & Yes & No & Yes \\ \hline
	\end{tabular}
	\label{tab:optics}
\end{table}
As we can see, the combination of the HRM, collimator and the top sawtooth lens gave the best combination of fringe visibility and sharpness of the central Bragg peak.
We note that since the sawtooth lenses were the farthest downstream from the X-ray source, these determined the extent of transverse coherence in the vertical direction (\emph{i.e.} along the $\unitvector{y}$ axis in the laboratory frame).
This was the configuration used in the HE-BCDI measurements throughout this paper.
 
\section{Partially coherent diffraction imaging}
\label{A:blurring}
The effect of partially coherent X-ray illumination on an isolated crystallite is modeled as a convolution of the ideal coherent diffraction pattern with a blurring function that reduces the visibility of the interference fringes~\cite{Tran2005,Clark2011,Clark2012}.
For transmission CDI measurements at present-day synchrotron sources, this blurring kernel is very effectively modeled as a multivariate Gaussian.
In this spirit we choose to model the three-dimensional partial coherence function as a multivariate Gaussian. 
If we consider the integer vector $N \equiv [ i~j~k ]^T$ to index the pixelated array space of the three-dimensional diffraction signal, then the blurring kernel expressed in this space is given by:
\begin{equation}\label{eq:blurinarray}
	g(N;C) = \frac{\sqrt{\det{C}}}{\left(2\pi\right)^{3/2}} \exp\left[-\frac{1}{2}N^T C N\right]
\end{equation}
Here $C$ is a $3 \times 3$ covariance matrix that we parameterize by its eigen-decomposition:
\begin{equation}
	C \equiv R\left(\psi, \unitvector{n}\right) \cdot D \cdot R^{-1}\left(\psi, \unitvector{n}\right)
\end{equation}
where $D$ is a (necessarily) non-negative diagonal matrix with elements $\{\lambda_1^2, \lambda_2^2, \lambda_3^2\}$ and $R(\psi, \unitvector{n})$ is a rotation matrix parameterized by the angle of rotation $\psi$ about an axis $\unitvector{n}$. 
We note that $\unitvector{n}$ in turn is parameterized by the spherical polar and azimuthal angles: $\unitvector{n} = \unitvector{n}(\theta, \phi)$. 
The partial coherence correction step in the image reconstruction process is essentially the following optimization problem over the entire set of $6$ blurring parameters $S \equiv \left\{\lambda_1, \lambda_2, \lambda_3, \psi, \theta, \phi\right\}$:
\begin{align}
S_\text{optimal} &= \arg\min_{S} \sum\limits_N I_\text{pc}(N) - I_\text{meas}(N) \log I_\text{pc}(N)	\label{eq.optimization} \\
	\text{where}~I_\text{pc}(N) &= g(N;C(S)) * I_\text{coh}(N) \notag
\end{align}
where $I_\text{meas}(N)$ is the measured partially coherent intensity, $I_\text{coh}(N)$ is the intermediate coherent diffraction intensity inferred from the phase retrieval iterations and $*$ denotes the convolution operation.
The objective function in Eq.~\eqref{eq.optimization} results from the maximum likelihood function of the Poisson-distributed intensity count of pixel $N$~\cite{Godard2012} and the optimization problem was solved using a simple gradient descent algorithm.

The phase retrieval recipe to recover the complex scattering object $\boldsymbol\rho(N)$ consisted of a combination of updates to an initial guess object using the error-reduction and hybrid input-output algorithms~\cite{Marchesini2007}. 
The exact sequence of iterations was as follows:
\begin{enumerate}
	\item 	$3750$ iterations of error reduction.
	\item	$150$ iterations of hybrid input-output.
	\item	$2750$ iterations of error reduction.
\end{enumerate}
A shrinkwrap algorithm~\cite{Marchesini2007} was employed after every $50$ iterations of error reduction to update the object support.
The update of the blurring kernel was initiated after iteration $1250$ of the first round of error reduction and was repeated after every $50$ iterations of error reduction thereafter, until iteration $2500$ of the secound round of error reduction.

We note that while more sophisticated methods of blind deconvolution have been applied to phase retrieval~\cite{Clark2012}, modeling the blurring kernel as a multivariate Gaussian has the advantage of being able to easily approximate the coherence length along any desired direction. 
We do this by first casting $g(N;C)$ from numerical array coordinates $N$ to 3D reciprocal-space coordinates $\boldsymbol{q}$ (with dimensions of inverse length):
\begin{equation}
    \boldsymbol{q} \equiv \mathbf{B}_q N
    \Longrightarrow N = \mathbf{B}_q^{-1} \boldsymbol{q}
\end{equation}
Here $\mathbf{B}_q$ is a $3 \times 3$ matrix whose columns denote the discrete sampling steps in reciprocal space.
 This ``basis'' of sampling vectors in 3D reciprocal space depends on the scattering and sample rotation  geometries and is typically determined in the course of generating the real-space image of the scatterer from the 3D complex array returned by the phase retrieval process.
Thus, we make a substitution for $N$ in Eq.~\eqref{eq:blurinarray} to get:
\begin{equation}
    g(\boldsymbol{q}; C) \sim \exp\left[-\frac{1}{2}\boldsymbol{q}^T \underbrace{\left(\mathbf{B}_q^{-T} C \mathbf{B}_q^{-1}\right)}_{\equiv C_q}\boldsymbol{q}\right]
\end{equation}
The coherence length $\ell_{\unitvector{v}}$ about any desired direction $\unitvector{v}$ is simply estimated by the standard deviation of the Gaussian in this direction:
\begin{equation}
    \ell_{\unitvector{v}} = \sqrt{\unitvector{v}^T C_q \unitvector{v}}
    \label{eq.cohlen}
\end{equation}
Table~\ref{tab:cohlengths} shows coherence lengths along specific directions computed by this method.
The values are given in terms of the standard deviation as well as the more conventional half-width of the Gaussian at the half-maximum value. 
We see from this table that the projected grain widths along these directions are greater than the estimated coherence lengths.

\begin{table*}
    \begin{tabular}{|c|c|c|c|c|}
    \hline
        \multicolumn{1}{|c|}{ $\hat{\boldsymbol{v}}$ } & 
        \multicolumn{1}{c|}{ \textbf{Description} } & 
        \multicolumn{2}{c|}{ \textbf{Coherence length (nm)} } & 
        \multicolumn{1}{c|}{ \textbf{Grain size (nm)} } \\
        \cline{3-4}
        \multicolumn{1}{|c|}{} & & \textbf{Standard deviation} & \textbf{Half-width at half max.} & \\
    \hline
        $\hat{\boldsymbol{x}}$ & Horizontal transverse & 293 & 345 & 464 \\ \hline
        $\hat{\boldsymbol{y}}$ & Vertical transverse & 306 & 360 & 406 \\ \hline
        $\hat{\mathbf{Q}}$ & Longitudinal & 285 & 336 & 477 \\ \hline
        $\hat{\boldsymbol{z}}$ & Hybrid & 238 & 280 & 323 \\ 
    \hline
    \end{tabular}
    \caption{
        Coherence lengths in different directions as estimated with Eq.~\eqref{eq.cohlen}.
        The width of the blurring kernel along the reciprocal lattice vector $\mathbf{Q}$ correspond to the longitudinal coherence lengths~\cite{Leake2009}.
    }
    \label{tab:cohlengths}
\end{table*}

\section{Strain computation}
\label{A:strain}
A coherent diffraction imaging measurement in the Bragg geometry is sensitive to the perturbations of the atomic positions over and above the average lattice periodicity, which may account for average strain over the volume of the crystalline scatterer as a whole. 
If the vector displacement of this perturbation is denoted by $\mathbf{u}(\boldsymbol{x})$ at a point $\boldsymbol{x}$ in the scatterer bulk, the the elastic strain tensor is given by:
\begin{equation}
	\mathcal{E}_{ij} = \frac{1}{2}\left(\partial_i u_j + \partial_j u_i\right)
	\label{eq.straindef}
\end{equation}
where the indices $i, j = 1, 2, 3$ and $\partial_i \equiv \partial/\partial x_i$ denotes the partial derivative with respect to the $i$th coorindate.
The component of the elastic strain in a direction denoted by a unit vector $\unitvector{n}$ is given by~\cite{Phillips2001}:
\begin{align}
	\mathcal{E}(\boldsymbol{x} \left| \unitvector{n} \right. )
		&= \sum\limits_i \sum\limits_j \hat{n}_i \mathcal{E}_{ij} \hat{n}_j  \label{eq.symmetric} \\
		&= \frac{1}{2} \sum\limits_i \sum\limits_j \hat{n}_i \left(\partial_i u_j\right)\hat{n}_j + \hat{n}_i \left(\partial_j u_i\right) \hat{n}_j \notag \\
		&= \frac{1}{2} \left[ \sum\limits_i \hat{n}_i \partial_i \left( \sum\limits_j u_j \hat{n}_j \right) + \sum\limits_j \hat{n}_j \partial_j \left( \sum\limits_i u_i \hat{n}_i \right) \right] \notag  \\
	\mathcal{E}(\boldsymbol{x} \left| \unitvector{n} \right.) &= \unitvector{n} \cdot \nabla \left[ \mathbf{u}(\boldsymbol{x}) \cdot \unitvector{n} \right] \label{eq.strainfinal}
\end{align}
If the reciprocal lattice point corresponding to the Bragg peak associated with a particular BCDI data set is given by $\boldsymbol{Q}$, then the measured phase from the phase retrieval process is given by: $\phi(\boldsymbol{x}) = 2\pi \mathbf{u}(\boldsymbol{x}) \cdot \boldsymbol{Q}$~\cite{Robinson2009}.
Here, $\boldsymbol{Q}$ has dimensions of inverse length \emph{i.e.} the strict reciprocal of $\mathbf{u}$.
Thus from Eq.~\eqref{eq.strainfinal}, the strain component along the reciprocal lattice vector $\boldsymbol{Q}$ is directly computed from the gradient phase:
\begin{equation}
	\boxed{
		\mathcal{E} (\boldsymbol{x} \left| \boldsymbol{Q} \right.) = \frac{\boldsymbol{Q}}{\left|\boldsymbol{Q}\right|}  \cdot \nabla \left(\frac{\phi(\boldsymbol{x})}{2\pi\left|\boldsymbol{Q}\right|}\right) }
	\label{eq.straincomp}
\end{equation}
Two things become evident from Eq.~\ref{eq.strainfinal}: 
\begin{enumerate}
    \item $\mathcal{E}( \boldsymbol{x} \left|\unitvector{n}\right.) = \mathcal{E}(\boldsymbol{x} \left|-\unitvector{n}\right.)$, \emph{i.e.} the computed strain is the same when the phase is measured along opposing reciprocal space directions (Friedel pairs).
    \item Finer resolution of the strain component is possible when computed from higher-order diffraction (\emph{i.e.} larger $\left|\boldsymbol{Q}\right|$). 
    This is seen by propagating the error in the measured phase into the strain: $\delta \epsilon \sim \nabla \left[\delta \phi \right]/ \left|\boldsymbol{Q}\right|$.
    Herein lies another significant advantage of performing BCDI at high beam energies: higher-order Bragg peaks become accessible due to their smaller scattering angles, which is not possible at lower X-ray energies.
    HE-BCDI therefore has the potential to deliver strain measurements at much higher resolution than low-energy BCDI.
\end{enumerate}
 
\section{Estimate of HE-BCDI spatial resolution}
\label{A:spatresol}
\begin{figure*}
    \centering
    \includegraphics[width=0.7\textwidth]{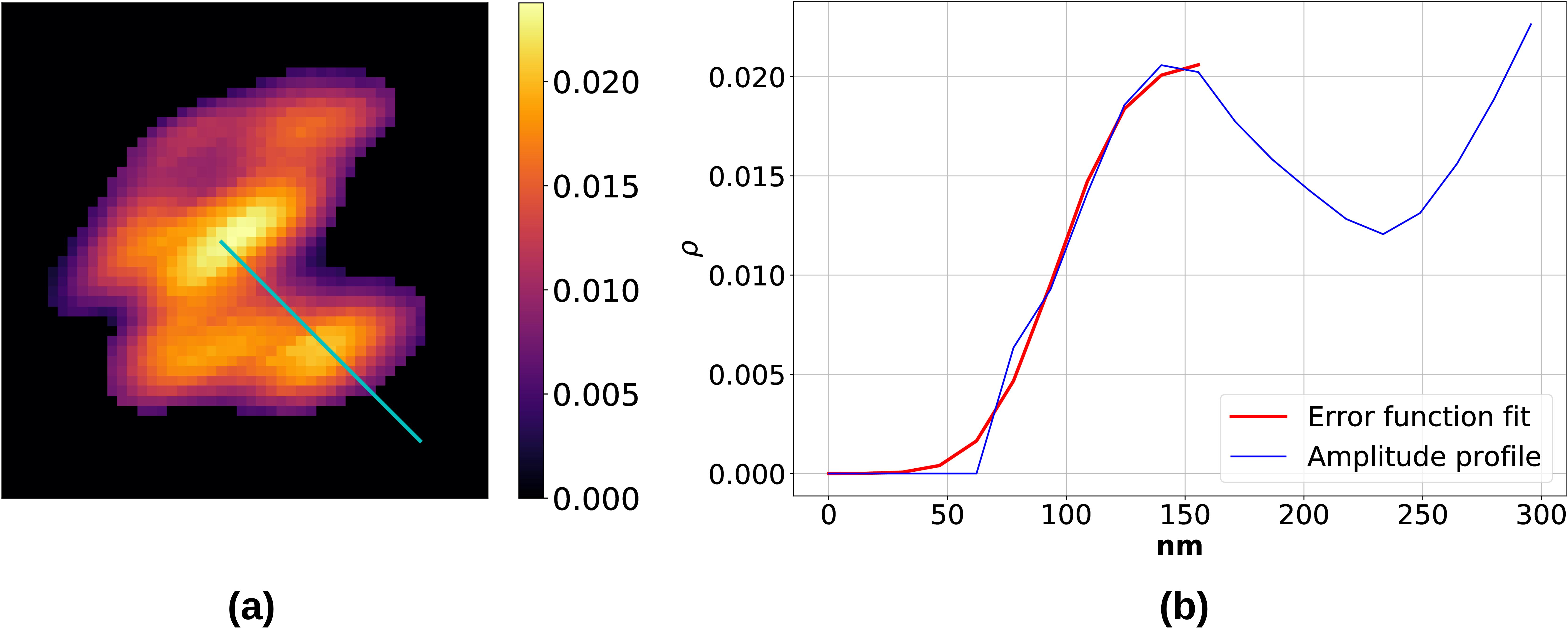}
    \caption{
        \textbf{(a)} The amplitude $\left|\boldsymbol{\rho}\right|$ of the polycrystal grain as determined from phase retrieval. 
        \textbf{(b)} variation of $\left|\boldsymbol{\rho}\right|$ along the line in (a) (from the exterior to the interior of the grain), along with the error function fit to the rising edge.
    }
    \label{fig:spatial}
\end{figure*}

An estimate of the spatial resolution of one of the grain reconstructions was made by fitting an error function to the line profile of the amplitude of the reconstructed grain (see Figure~\ref{fig:spatial}).
The error function width was determined to be $47.5$ nm, which is an estimate of the spatial resolution of the measurement. 

\section{HE-BCDI from a $\left\langle 200 \right\rangle$ Bragg peak}
\label{A:highorddiff}
\begin{figure}
    \centering
    \includegraphics[width=0.5\textwidth]{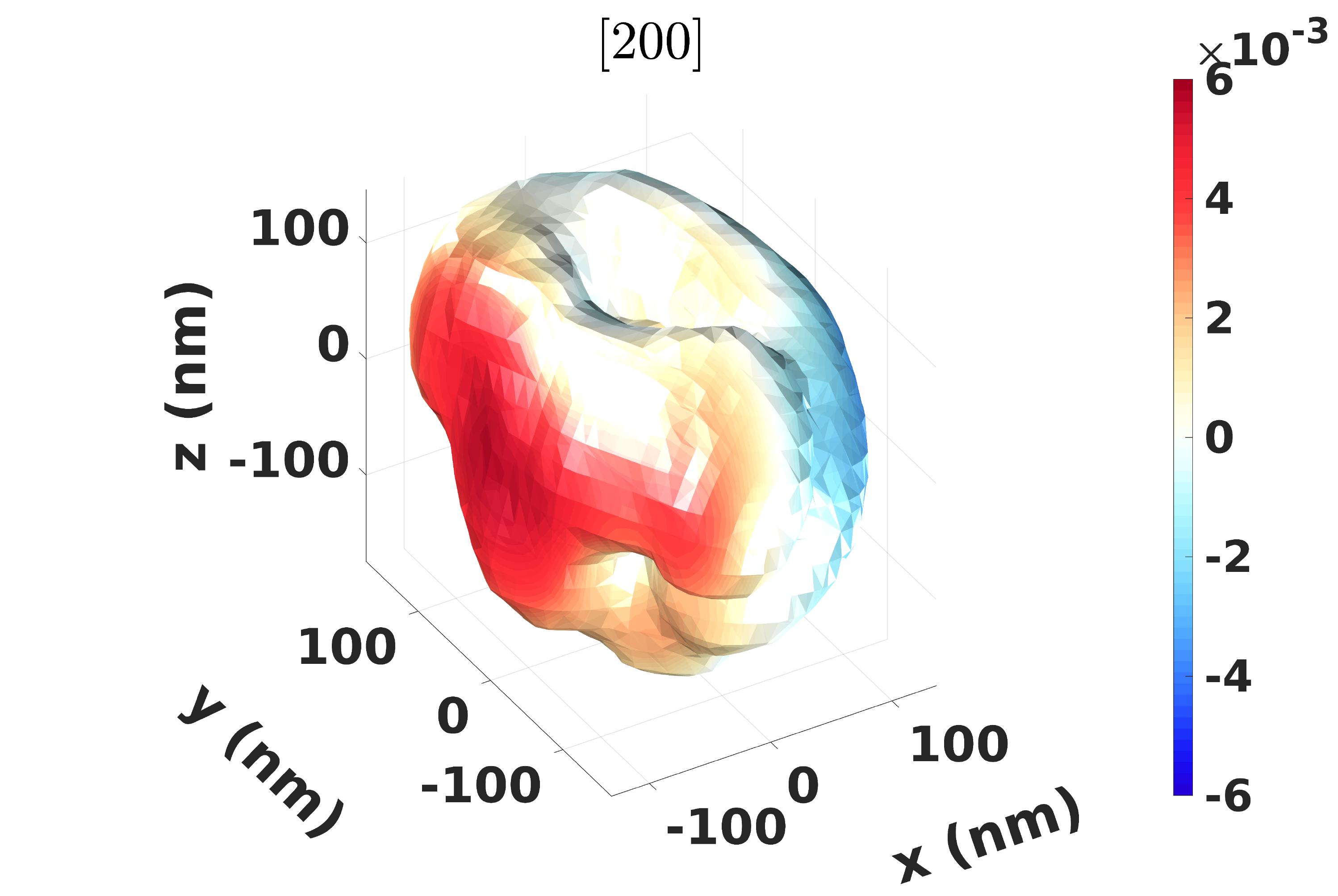}
    \caption{
    Spatially resolved strain component along the surface of the smaller of the two chosen grains.
    }
    \label{fig:200grain}
\end{figure}

Figure~\ref{fig:200grain} shows the HE-BCDI reconstruction of another, smaller grain in the polycrystalline film. 
The partially coherent diffraction data was obtained at a very manageable scattering angle of $6.7^\circ$.

%

\end{document}